\documentclass[onecolumn,12pt]{IEEEtran}

\usepackage{algorithm,algorithmic,amsbsy,amsmath,amssymb,epsfig}

\DeclareMathAlphabet{\mathpzc}{OT1}{pzc}{m}{it}
\usepackage{times,amsmath,epsfig}
\usepackage{cite}
\usepackage{graphicx}
\usepackage{psfrag}
\usepackage{subfigure}
\usepackage{url}
\usepackage{stfloats}
\usepackage{amsmath}
\usepackage{amssymb}
\usepackage{array}
\usepackage{color,soul}
\usepackage{epstopdf}
\usepackage{enumerate}

\newcommand{\beq}{\begin{equation}}
\newcommand{\eeq}{\end{equation}}
\newcommand{\bitm}{\begin{itemize}}
\newcommand{\ba}{\begin{array}}
\newcommand{\ea}{\end{array}}
\newcommand{\eitm}{\end{itemize}}
\newcommand{\beqn}{\begin{eqnarray}}
\newcommand{\eeqn}{\end{eqnarray}}
\newcommand{\beqno}{\begin{eqnarray*}}
\newcommand{\eeqno}{\end{eqnarray*}}
\newcommand{\bma}{\begin{displaymath}}
\newcommand{\ema}{\end{displaymath}}
\newcommand{\bnu}{\begin{enumerate}}
\newcommand{\enu}{\end{enumerate}}
\newcommand{\bce}{\begin{center}}
\newcommand{\ece}{\end{center}}
\newcommand{\btb}{\begin{tabular}}
\newcommand{\etb}{\end{tabular}}

\begin{document}

\title{Full-Duplex Cognitive Radio: A New Design Paradigm for Enhancing Spectrum Usage}

\author{\IEEEauthorblockN{Yun Liao\IEEEauthorrefmark{1}, Lingyang Song\IEEEauthorrefmark{1}, Zhu~Han\IEEEauthorrefmark{2}, and Yonghui Li\IEEEauthorrefmark{3}\vspace{3mm}\\}
\IEEEauthorblockA{\IEEEauthorrefmark{1}State Key Laboratory of Advanced Optical Communication Systems and Networks,\\School of Electronics Engineering and Computer Science, Peking University, China.
\\ \IEEEauthorrefmark{2} Electrical and Computer Engineering Department, University of Houston, USA.
\\ \IEEEauthorrefmark{3}School of Electrical and Information Engineering, The University of Sydney, Australia.
}}

\maketitle

\thispagestyle{empty}
\begin{abstract}
With the rapid growth of demand for ever-increasing data rate, spectrum resources have become more and more scarce. As a promising technique to increase the efficiency of the spectrum utilization, cognitive radio~(CR) technique has the great potential to meet such a requirement by allowing un-licensed users to coexist in licensed bands. In conventional CR systems, the spectrum sensing is performed at the beginning of each time slot before the data transmission. This unfortunately results in two major problems: 1) transmission time reduction due to sensing, and 2) sensing accuracy impairment due to data transmission. To tackle these problems, in this paper we present a new design paradigm for future CR by exploring the full-duplex~(FD) techniques to achieve the simultaneous spectrum sensing and data transmission. With FD radios equipped at the secondary users~(SUs), SUs can simultaneously sense and access the vacant spectrum, and thus, significantly improve sensing performances and meanwhile increase data transmission efficiency. The aim of this article is to transform the promising conceptual framework into the practical wireless network design by addressing a diverse set of challenges such as protocol design and theoretical analysis. Several application scenarios with FD enabled CR are elaborated, and key open research directions and novel algorithms in these systems are discussed.
\end{abstract}


\newpage

\section{Introduction}

The existing and new wireless technologies, such as smart phones, wireless computers, WiFi home and business networks are rapidly consuming radio spectrum. Unlike the wired Internet, the wireless world has a limited amount of links to distribute. Consequently, the traditional regulation of spectrum requires a fundamental reform in order to allow for more efficient  use of the precious resource of the airwaves~\cite{Song2010}. Cognitive radio (CR) has been widely recognized as a promising technique to increase the efficiency of spectrum utilization~\cite{mitola1999cognitive}. It allows un-licensed secondary users~(SUs) to coexist with primary users (PUs) in licensed bands. There are basically two kinds of CR networks~(CRNs): underlay and overlay. In underlay CRNs, SUs transmit at a limited power such that they do not cause harmful interference to the primary network. The restricted power limits the transmit range of SUs. In overlay mode, SUs need to accurately sense the transmission of PUs' spectrum, identify unused spectral holes to transmit and leave it when the incumbent radio system is ready to transmit~\cite{yucek2009survey}. In this article, we concentrate on overlay CRNs.

Most existing CRNs deploy half-duplex~(HD) radios to transmit and receive signals in two orthogonal channels and SUs employ the well-known ``Listen-before-Talk''~(LBT) protocol, in which SUs sense the target channel before transmission~\cite{yucek2009survey}. Though proved to be effective, the LBT protocol actually dissipates the precious resources by employing a time-division duplexing, and thus, unavoidably suffers from two major problems:
\begin{enumerate}
  \item The SUs have to sacrifice transmission time for spectrum sensing, and even if the spectrum hole is long and continuous, the data transmission needs to be split into small discontinuous slots;
  \item During SUs' transmission, SUs cannot detect the changes of PUs' states, which leads to collision when PUs arrive and the spectrum waste when PUs leave.
\end{enumerate}

Hence, it would be desirable for SUs to continuously sense spectrum and meanwhile transmit whenever spectrum holes are detected. This, however, seems impossible with the conventional HD systems. A full-duplex~(FD) system, where a node can send and receive signals at the same time and frequency resources, offers the potential to achieve simultaneous sensing and transmission in CRNs. However, for many years, it was considered impractical~\cite{Choi2010} because signal leakage from local output to input, referred to as self interference, may overwhelm a receiver, and thus makes it impossible to extract the desired signal. However, with the recent significant process in self interference cancelation, it has presented a great promise to realize  FD communications for future wireless networks~\cite{Choi2010}.

In this article, we explore the FD techniques in CRNs, and present a novel ``Listen-and-Talk''~(LAT) protocol, by which the SUs can simultaneously perform spectrum sensing and data transmission~\cite{liao2014gc}. Specifically, by equipping with FD radios, SUs can sense the target spectrum band in each time slot, and determine if the PUs are busy or idle, and at the same time they can also transmit data or decide to keep silent based on teh sensing results. Apparently, the proposed FD-CR system is totally different from the traditional HD based one in many aspects, including:
\begin{itemize}
  \item Spectrum sensing: in FD-CR, sensing is continuous, but the received signal for sensing is interfered by the residual self-interference~(RSI), which degrades the signal-to-interference-plus-noise ratio~(SINR) in sensing. While in HD-CR, there exists no RSI in received signal for sensing, but the sensing process is discontinuous and only takes a small fraction of each slot. This leads to unreliable sensing performance due to the inefficient number of samples to make decisions;
  \item Data transmission: in traditional HD-CR, the SUs can only utilize the remaining part of each slot after sensing for data transmission. On contrary, in FD-CR, SUs can continuously transmit as long as PUs are absent. However, in FD-CR, the data transmission affects the sensing process, and thus there exists a constraint of transmit power to achieve acceptable sensing performance.
\end{itemize}

In summary, the FD technology enables to explore another dimension of the network resources for increasing the capacity of CRNs. This thus requires a new design of signal processing techniques, resource allocation algorithms, and network protocols. For example, one of the major challenges faced by FD-CR is how to optimize the transmit power for the FD source node to maximize the system throughput. This article comprehensively discusses the novel protocol design issues, key system parameter derivation, and practical algorithms for FD-CR systems. In addition, we also extend the proposed FD-CR to {\em distributed} and { \em centralized} network scenarios, and demonstrate the new research challenges and provide the latest promising research development. Some potential research directions and open problems will be also discussed.

\section{Basics of Full-Duplex Communications and Cognitive Radio}

This section briefly introduces a basic FD communication system and the conventional LBT protocol.

\subsection{Full-Duplex Communication System}

\begin{figure}[ht]
\centering
\includegraphics[width=4in]{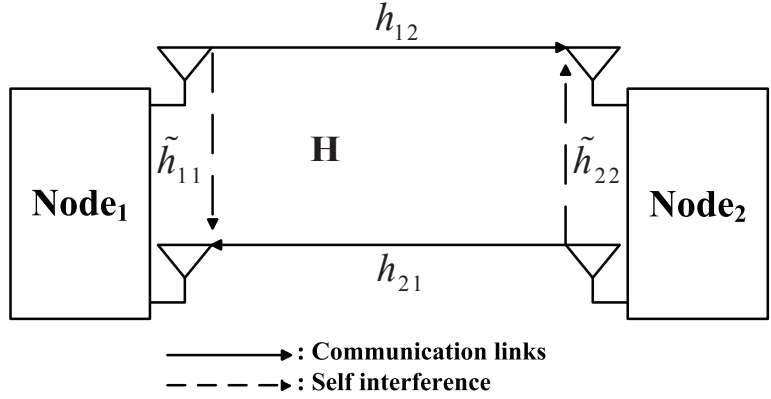}
\caption{Full-Duplex Wireless Communications} \label{fd}
\end{figure}

Fig.~\ref{fd} presents a simple two-node FD communication system with a transmit and a receive antenna at each node. The FD technique allows two nodes to transmit and receive the signals at the same frequency and time interval, i.e., node $i$ can transmit its signal $x_i~\left(i=1,2\right)$ with the transmit antenna and receive $x_{3-i}$ from the other node via the receive antenna in the same channel simultaneously. However, this leads to severe self-interference caused by the signal leakage from the transmit RF unit to the receive RF unit, as shown in Fig.~\ref{fd}.
In practical FD systems, the self-interference cannot be completely canceled, such that the signals received at each node is a combination of the signal transmitted by the other source, the RSI, and the noise. Specifically, the RSI can be typically modeled as AWGN, Rayleigh or Rician distributed variables~\cite{Choi2010,zhou2014}, of which the variance is proportional to the transmit power.

\subsection{Conventional Cognitive Radio Communication System}

In a conventional CRN, SUs are allowed to access the spectrum allocated to PUs without causing severe interference at the primary network. The typical assumptions are that PUs do not notify SUs when they begin or stop transmission, and do not wait when SUs are using the channel. Thus, SUs need to reliably identify the spectrum holes to access and backoff when PUs appear. Typically, SUs' traffics are slotted, and every SU's time slot is divided into two sub-slots for sensing and data transmission~\cite{yucek2009survey}. At the beginning of each slot, all SUs sense the spectrum with a certain sensing strategy such as energy detection, matched filter detection, cyclostationary feature detection, or cooperative detection, and test the following two hypothesises: (1) the spectrum is idle; and (2) the spectrum is occupied by a PU. Once the spectrum is judged idle, some SUs transmit in the rest time of the slot according to a certain scheduling scheme to minimize the probability of potential collision among SUs~\cite{Xin2008}.


\begin{figure}[ht]
\centering
\includegraphics[width=7in]{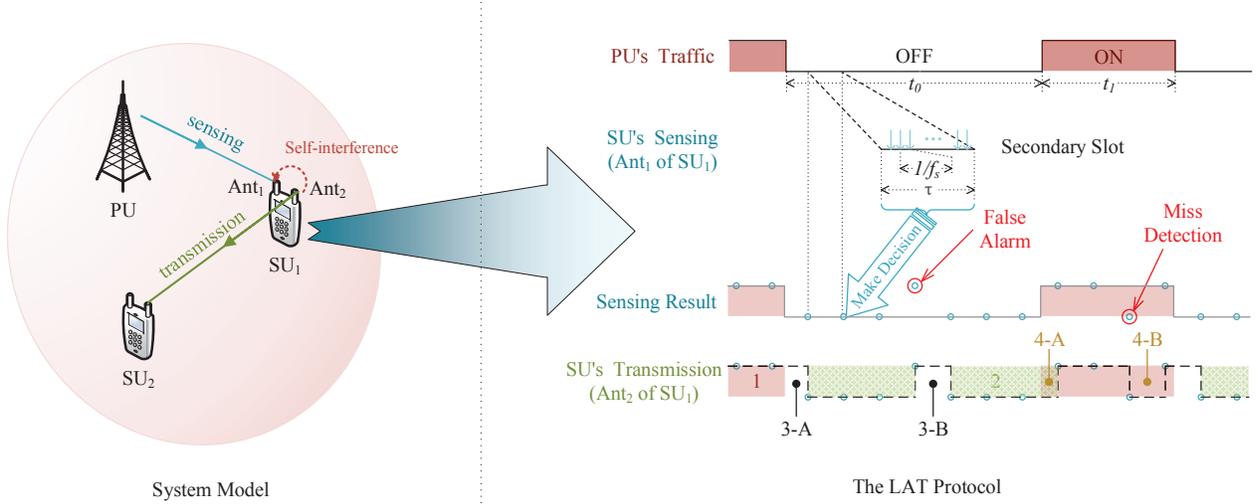}
\caption{System Model of the LAT Protocol.} \label{LAT}
\end{figure}
\section{Listen-and-Talk~(LAT) Protocol}

In this section, we present our proposed LAT protocol by exploring the FD techniques at SUs in CRNs.

\subsection{The LAT Protocol}

Fig.~\ref{LAT} shows the system model of the LAT protocol. Consider a CR system consisting of one PU and one SU pair~(see the left side of Fig.~\ref{LAT}), in which SU$_1$ is the secondary transmitter and SU$_2$ is the receiver. Each SU is equipped with two FD antennas Ant$_1$ and Ant$_2$. SU$_1$ performs sensing and transmission simultaneously: Ant$_1$ senses the spectrum continuously while Ant$_2$ transmits data when a spectrum hole is detected. Specifically, at the end of each slot, SU$_1$ makes the decision on the PU's presence, and determines whether to transmit or not in the next slot. Assume that the PU can utilize the spectrum freely, and thus, there exist the following four cases of spectrum utilization, as shown in the right side of Fig.~\ref{LAT}:
\begin{itemize}
    \item Case$_1$: the spectrum is occupied only by the PU, and SU$_1$ is silent.
    \item Case$_2$: the PU is absent, and SU$_1$ utilizes the spectrum.
    \item Case$_3$: neither the PU nor SU$_1$ is active, and there remains a spectrum hole.
    \item Case$_4$: the PU and SU$_1$ both transmit, and a collision happens.
\end{itemize}

Among these four cases, Case$_1$ and Case$_2$ are the normal cases, and Case$_3$ and Case$_4$ are referred to as spectrum waste and collision, respectively. The spectrum waste ratio and the collision ratio are the most important metrics to measure sensing performance. One key challenge in using the FD technique is that the transmit signal at Ant$_2$ of SU$_1$ is received by Ant$_1$, causing irreducible RSI, which then has strong impacts on the sensing performance. Specifically, when Ant$_2$ is idle, the received signal at Ant$_1$ only contains the potential PU's signal and noise, and is free of RSI; once the data transmission begins, the RSI is introduced to the received signal, which may lead to a sharp decrease of SINR in sensing.

\subsection{Key Design Parameters}

The protocol design needs to satisfy the strict constraint that the interference of SUs to the primary network should not exceed a certain level of the probability of collision:
\begin{equation}\label{Pc}
P_c = \lim_{t\rightarrow\infty}\frac{\text{Collision duration}}{\text{PU's transmission time}}.
\end{equation}
From \eqref{Pc}, the sensing strategy, thresholds, and slot length can be derived and optimized as shown in Table~\ref{KPD}.

\begin{table}[t]
\renewcommand{\arraystretch}{1.2}
\caption{Key Design Parameters in the LAT} \label{KPD}
\centering
\begin{tabular}{|p{3cm}|p{13cm}|}
\hline
\centering{\textbf{Key Parameters}} & {\textbf{Design Methods}} \\ \hline
Sensing Scheme & The LAT protocol uses energy detection as the sensing scheme, which requires less priori information of PU's signal patterns than other detection strategies\footnotemark; however, the impact of RSI needs to be carefully considered since the average received power increases significantly with the transmit power.\\ \hline
Sensing Threshold &  The expected received power varies with the transmission of SUs due to the RSI, and the expected received power increases with the transmit power. Thus, the sensing threshold at a certain SU under the LAT protocol is no longer fixed, but adjusted with its transmit power. Thus, the SU needs to check its own activity and choose the best threshold accordingly in each slot.\\ \hline
Length of Time Slot& Generally, a longer time slot length leads to better sensing performances since more samples are considered, but, the probability that the PU changes its state in one slot also increases, which would make the sensing unreliable. Thus, a modest slot length needs to be carefully designed.  \\ \hline
\end{tabular}
\end{table}

\footnotetext{Note that other detection methods can also be used in the sensing process\cite{riihonen2014}.}

\subsection{Performance Analysis}
\subsubsection{Power-throughput Tradeoff}

\begin{figure}[ht]
\centering
\includegraphics[width=5in]{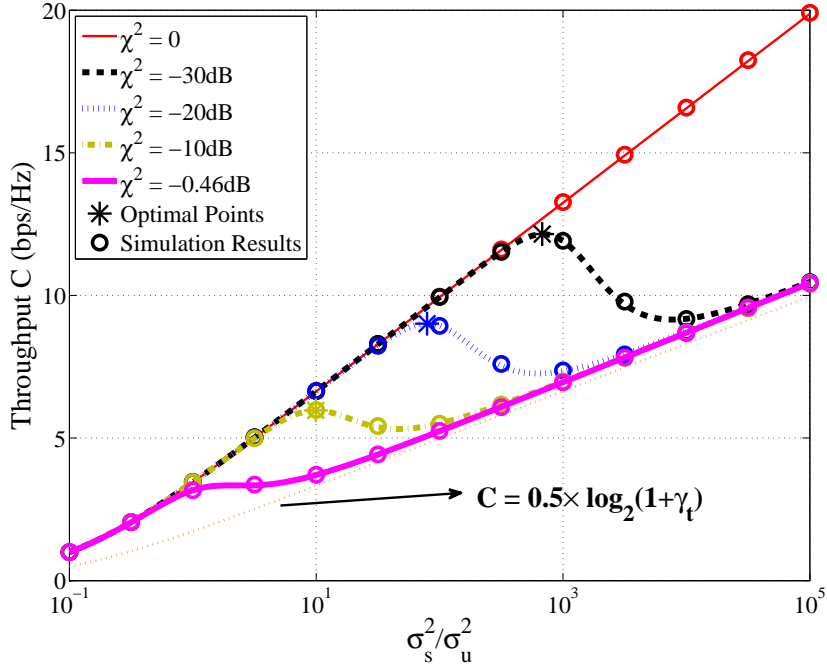}
\caption{Power-Throughput Curves, in which the curves represent the analytical performance, and the circles on curves show the simulation results.} \label{power_thput}
\end{figure}

The saturated throughput of SUs can be calculated as
\begin{equation}\label{r_lat}
C = R \cdot \left( {1 - P_{w}} \right),
\end{equation}
where $P_{w}$ represents the ratio of wasted spectrum hole and $R$ denotes the transmission rate of the channel. Let's consider the relation between transmit power and the two factors $R$ and $\left(1-P_{w}\right)$. On one hand, the rate $R$ is positively proportional to SU's transmit power. On the other hand, the sensing performance degrades with the increase of RSI, i.e., with the same RSI level, the false alarm probability increases with the transmit power under the same constraint of the collision probability\cite{liao2014gc}. And this, in turn, leads to the lift of $P_w$ and the decrease of $\left(1-P_{w}\right)$.

Thus, there exists a power-throughput tradeoff in this protocol: when the secondary transmit power is low, the RSI is negligible, and the spectrum will be used more efficiently, yet the ceiling throughput is limited by $R$; when the transmit power increases, the channel rate becomes higher, but the sensing performance is degraded due to the RSI.

Fig.~\ref{power_thput} shows the throughput performance of the LAT protocol in terms of secondary transmit power. The small circles are the simulated results and the thin solid line depicts the ideal case with perfect RSI cancelation. Without the RSI, the sensing performance is not affected by transmit power, and thus the throughput always goes up with the power. It can be used to serve as an upper bound for the LAT performance. The thick dash-dotted, dotted and dash lines in the middle are the typical cases, in which we can clearly observe the power-throughput tradeoff and identify the local optimal power~(the asterisks in the figure). With the decrease of the RSI~($\chi^2=\frac{\text{power of RSI}}{\text{SU's transmit power}}$ for -10dB, -20dB, and -30dB), the optimal transmit power increases, and the corresponding throughput goes to a higher level. This is because the smaller the RSI is, the closer it approaches the ideal case, and the deterioration cause by self-interference becomes dominant under a higher power. However, when $\chi^2$ is sufficiently large and close to 1, there exists no power-throughput tradeoff~(see the thick solid line denoting the cases when $\chi^2 = -0.46\text{dB} = 0.9$). The sensing performance is unreliable even if the transmit power is small and no local optimal point can be found in this curve. One noticeable feature of the LAT protocol shows in Fig.~\ref{power_thput} is that when self-interference exists, all curves approach the thin dotted line $C = 0.5R$ when the power goes up. This line indicates the case that the spectrum waste is 0.5, i.e., the sensing is totally ineffective. When the transmit power is too large, severe self-interference largely degrades the performance of spectrum sensing. And the larger $\chi^2$ is, the earlier the sensing gets unbearable and the throughput approaches the line for $C=0.5\times \log_2(1+\gamma_t)$.

\begin{table}[t]
\renewcommand{\arraystretch}{1.3}
\caption{Comparison between LAT and LBT} \label{Comparison}
\centering
\begin{tabular}{|p{4cm}<{\centering}|p{6.5cm}<{\centering}|p{6.5cm}<{\centering}|}
\hline
\centering{Parameters} & LAT & LBT \\ \hline
Spatial Correlation & \centering{NA} & Both sensing and transmission deteriorate with the increase of spatial correlation. \\ \hline
Sensing Duration & \centering{NA} & Non-monotonous impact. An optimal ratio of sensing duration can be found\cite{liang2008twc}. \\ \hline
Secondary Slot Length & Only sensing performance needs to be considered in the optimization. The optimal slot length is typically similar to sensing duration in LBT. & It requires the joint optimization of the durations of sensing and transmission sub-slots. \\ \hline
Secondary Transmit Power & The power-throughput tradeoff exists, and a local optimal transmit power can be found. & Throughput increases with the power monotonously. \\ \hline
RSI Level & Better suppression level leads to higher local optimal power and throughput. & NA\\
 \hline
\end{tabular}
\end{table}

As shown in both analytical and simulated results, in low power region, the SUs' throughput first increases and then decreases such that a local optimal transmit power can be found; while in high power region, the SUs' throughput increases monotonically with the power. For each SU, the maximum transmit power is limited by the physical structure. The existence of the local optimal transmit power indicates that the maximum power may not always be the best choice to achieve the highest throughput. Instead, if the maximum power exceeds the local optimal one, yet it is not high enough, a modest power may be much better than the maximum one in achieving both higher throughput and longer transmission time.

\begin{figure}[h]
\centering
\includegraphics[width=4.2in]{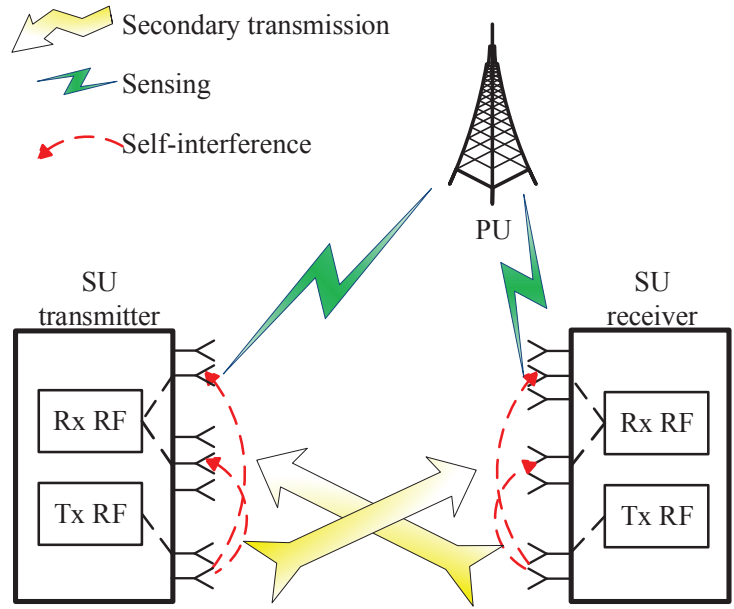}
\caption{Full-Duplex Cognitive MIMO System}
\label{mimo}
\end{figure}

\subsubsection{Comparison between the LAT and Conventional LBT Protocol}

Since there exist limitations for both LBT and LAT protocols, we make comparisons under the same model shown in Fig.~\ref{LAT}. For fairness, in the LBT, SU$_1$ uses both antennas to sense and transmit, and correlation between the two antennas needs to be considered. In the LBT, the data transmission time is reduced because of spectrum sensing, while in the LAT, the RSI is the main problem that decreases the performance\footnote{Note that if the LAT system is compared with conventional LBT system consisting of SUs with single antenna, the comparison would be more straightforward since there exists no spatial correlation in the LBT. Also, the switching would still exist due to the effects of RSI.}. Here we list some important parameters that influence the SUs' throughput in Table~\ref{Comparison}. Note that the proposed LAT may not always outperform the conventional protocol, especially when secondary transmit power is sufficiently large and the impact of RSI becomes unbearable. However in most cases, the LAT performs better due to its higher utilization efficiency of the spectrum holes and prompt detection and reaction to the PU's state change.

\section{Full-Duplex Cognitive Radio Applications}

In this section, we focus on some key unique applications of FD CR systems.
\subsection{Distributed Scenarios}

The distributed scheme does not have any central infrastructure for coordinating the common channel access procedure. Hence, each FD-CR user that is going to transmit has to take the contention procedures and resolve possible collisions.

\subsubsection{Distributed Spectrum Access Scheme}
We first introduce a simple distributed FD-CR system. Consider $M$ non-cooperative FD-SUs contending for one channel of PUs~\cite{liao2015icc}. Each SU senses the channel by the LAT protocol independently, and accesses the spectrum without communicating with each other. To avoid collision among SUs, a distributed dynamic spectrum access~(DSA) scheme is needed. The difference of using the FD technique in the design of DSA scheme is that PUs are no longer ``blind'' to SUs when SUs are transmitting data, instead, SUs can detect in real time the state changes of PUs and other SUs continuously, and when collision happens, SUs can backoff immediately before finishing the current packet. This brings about the possibility of a new DSA scheme with the less spectrum waste and shorter collision length.

\subsubsection{Full-Duplex MIMO System}
 As shown in Fig.~\ref{mimo}, it consists of a pair of FD MIMO transceivers, nodes $A$ and $B$ , where each node is equipped with multiple antennas~($N$), respectively. In each node, some antennas~($N_s$) are used for sensing, some~($N_t$) for data transmission, and some~($N_r$) can be used for receiving data from the other CR node. Both nodes operate in the same frequency band at the same time. Hence, if $N_s = N$, the system becomes the traditional CR with LBT; When all these three parameters are employed, this system supports bi-directional communication while sensing, but the interference is quite complicated among the antennas.

\subsection{Centralized Scenarios}
In the centralized scenarios, an access point~(AP) establishes the connection with the mobile users, which are served in the coverage area.
\subsubsection{Full-Duplex Cognitive Access Point System}

\begin{figure}[ht]
\begin{center}
\centerline{\includegraphics[width=5in]{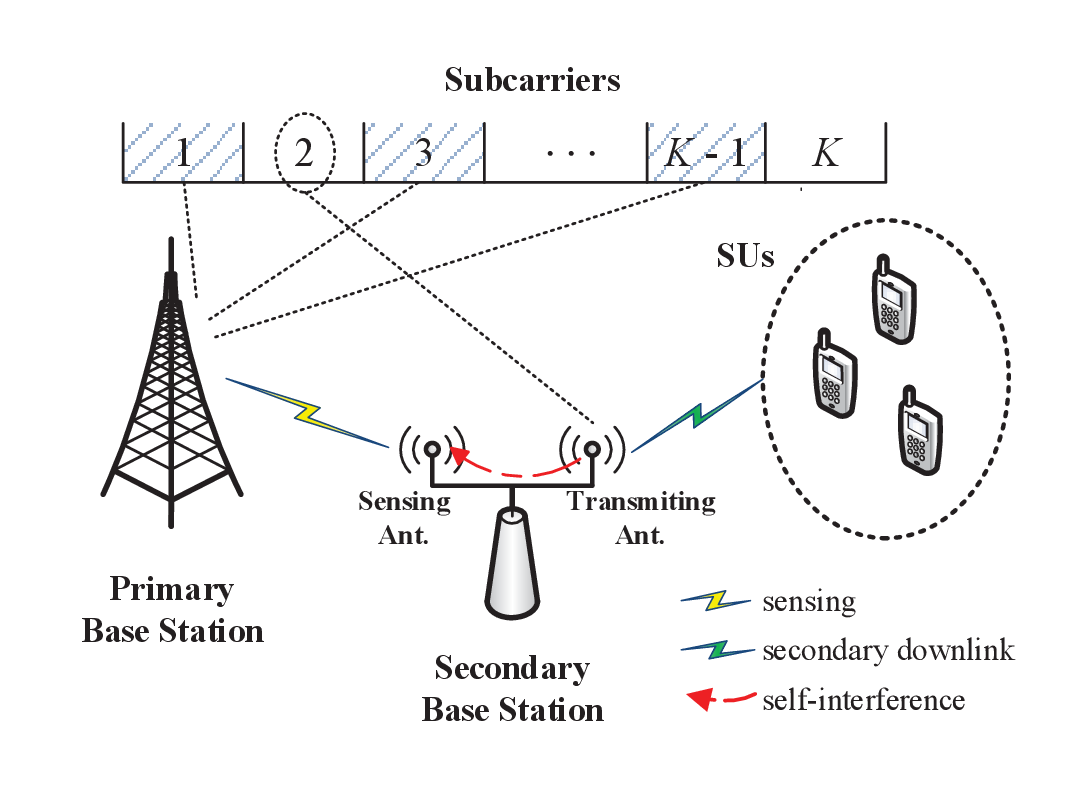}}
\caption{Full-Duplex Cognitive Access Point System} \label{ofdma}
\end{center}
\end{figure}

Fig.~\ref{ofdma} shows a simple secondary AP system consisting of a FD cognitive OFDM-AP with 2 antennas and a number of SUs~\cite{Wang2015icc}. One antenna in the AP senses the PU's spectrum, and the other antenna provides wireless service for the SUs, in which the downlink transmission brings the RSI to the sensing results. Hence, how to assign SUs to the proper subcarriers and adjust the transmit power becomes essential for the system performance.

\subsubsection{Full-Duplex Cognitive Relay System}

\begin{figure}[h!]
\centering
\includegraphics[width=4in]{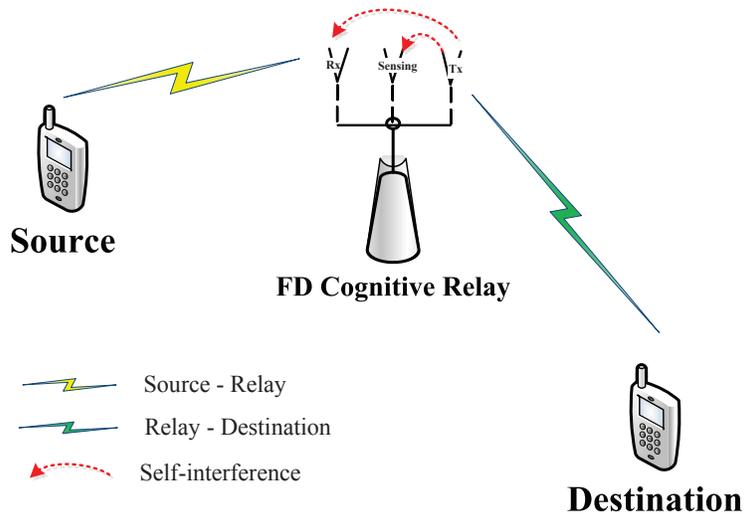}
\caption{{Full-Duplex Cognitive Relay System.}}\label{relay}
\end{figure}

By equipping the relay node with a FD radio, the relay can receive and retransmit signals simultaneously, and thus spectral efficiency can be improved, compared to the traditional HD relay. Fig.~\ref{relay} illustrates a simple FD cognitive relay system consisting of one source, one destination, and one FD relay node. Both the source and relay nodes use the same time-frequency resource and the relay node works in the FD mode with 3 antennas (one for spectrum sensing, one for reception, and one for transmission). The communication process can be briefly described below:
\begin{itemize}
  \item The source transmits signals to the FD relay;
  \item At the same time, the FD cognitive relay performs spectrum sensing;
  \item And the relay forwards the signals received in the previous time slots to the destination.
\end{itemize}
Note that different from traditional HD relay, the FD relay uses two antennas to receive data from the PU for sensing and the source for data forwarding. Thus, since FD works in the same frequency band, these two receive antennas actually have a combination of the PU, source, and RSI signals.

\section{Research Problems}

In this section, we summarize the main research problems for FD-CR communication systems. Similar to the traditional wireless systems, multi-dimensional resource on space, time, frequency, power, and nodes need to be properly managed to optimize the overall system performance. Specially, FD communication provides another dimension of resource and its performance is also greatly affected by the RSI. Some key research problems as well as possible solutions are also summarized.

\subsection{Signal Processing Techniques}

\subsubsection{Spectrum Sensing}
The non-cooperative narrowband sensing has been elaborated in Section~III. However, the degradation of sensing performance becomes unbearable when transmit power is high. A promising solution for this problem is \emph{cooperative sensing}, in which the sensing results of several SUs are combined to make a final decision. However, when employing cooperative spectrum sensing in FD-CRNs, there is interference from the transmitting SU to other SUs, which makes the local sensing different from the conventional HD case\cite{liao2014iccs}.

Also, since CRNs will eventually be required to exploit spectrum opportunities over a wide frequency range, a FD-enabled \emph{wideband sensing scheme} is needed. With the impact of RSI, the original sparsity, which is the base for conventional wideband sensing scheme, would change, and the whole sensing scheme may be different.

\subsubsection{Multiple Antenna Techniques}

If multiple antennas are equipped at the FD-CRs, beamforming and antenna selection can be employed to further improve the secondary network performance:

\begin{itemize}
  \item Transmit beamforming: transmit beamforming is used to control the directionality of transmission in order to provide a large antenna array gain in the desired directions. For a FD cognitive MIMO system, the transmit antenna set at each FD-CR node can perform transmit beamforming to simultaneously transmit information and reduce the interference to its own received sensing signals. The design is to jointly optimize the sum rate of the system. If the FD cognitive AP node, which serves a group of users, is equipped with multiple antennas, it may be able to support multiple downlink transmissions while maintaining reliable sensing performance by using certain structure of antennas to minimize the RSI.

  \item Antenna selection: for a FD-CR node, especially a node with multiple antennas, each antenna can be configured to sense (receive) or transmit the signals. This creates an important problem to optimally select the antenna configurations optimize the system performance~\cite{zhou2014}. In a FD cognitive MIMO system, the problem is to choose one group of antennas to sense the spectrum, one group to transmit and the rest to receive signals from another SU simultaneously. Such a combinatorial problem becomes much complicated as the number of antennas increases. Similarly, in a general FD cognitive relay system, each relay can adaptively select its sense and transmit and receive antennas based on the instantaneous channel conditions to achieve reliable sensing as well as maximum SINR in transmissions.
\end{itemize}

\subsection{Dynamic Spectrum Access and Management}

DSA is the key approach in the CRNs, through which, a cognitive wireless node is able to adaptively and dynamically transmit and receive data in a changing radio environment.

\subsubsection{Distributed Dynamic Spectrum Access}

In many scenarios such as in ad hoc CRNs where the SUs compete for several PU channels, deploying a central controller is not always possible. Therefore, distributed DSA will be required, by which each SU has to independently gather, exchange, and process the information of the wireless environment. The commonly used CSMA/CA in distributed DSA with HD users can effectively reduce collision probability~\cite{Xin2008}, but some problems still exist: (1) collision among the SUs can never be detected if the SUs are synchronized, such that the secondary transmission may fail to a large scale, and (2) SUs cannot abort transmission when collision happens, which leads to long collision duration. With FD-CR, the SUs can not only detect the presence of PUs, but also detect collision with other SUs during transmission, such that the collision duration is reduced significantly. But the RSI may degrade the collision detection accuracy, which cannot be ignored~\cite{liao2015icc}.

\subsubsection{Centralized Dynamic Spectrum Access}

For the centralized DSA, a central controller is deployed to gather and process information about the wireless environment. Therefore, the decision of the SUs to access the spectrum can be made such that the desired system-wide objectives are achieved. This centralized control method fits very well the FD cognitive AP/relay scenarios, where the AP/relay performs spectrum sensing and monitors the users below. Specifically, in a FD cognitive AP system consisting of one FD OFDM AP and some downlink SUs, a fundamental challenge is how to match available subcarriers with downlink SUs to optimize the network performance~\cite{Wang2015icc}. In this scenario, the consideration of tolerable RSI at the AP for reliable sensing is the main difference from conventional radio allocation problems. Similarly, in a FD cognitive relay system consisting of multiple secondary source and destination pairs and a FD relay node, the corresponding subcarriers should be properly allocated taking account of the RSI.

\subsubsection{Power Control}

Power control has been a commonly used approach in multi-user communication systems to optimize system performance such as link data rate, network capacity and coverage. Unlike traditional wireless networks, FD-CR communication suffers from RSI which deteriorate sensing performance when transmit power increases. Therefore, the corresponding power control algorithm needs to be properly redesigned in order to optimize system performance. Different power constraints, e.g. {\em total} or {\em individual} transmit power, will lead to different designs and final solutions. Moreover, as detailed below, different FD-CR systems require different power control algorithms:

\begin{itemize}
  \item FD cognitive MIMO system: the antennas at the FD node are divided into sensing, receive and transmit antenna sets with individual power constraints. Considering the RSI at the sensing set, the optimal power pouring mechanism can be significantly different from the conventional water-filling.
  \item FD cognitive relay system: the relay is under individual power constraint, and both the relayed signals and sensing results are corrupted by the RSI. Increasing the transmit power at the relay will increase the SNR at the destination, but on the other side decrease the accuracy of sensing and blur the received signals from the source. The optimization needs to consider these issues.
  \item FD cognitive AP system: the FD cognitive AP should sense the spectrum while communicating with the SUs at the same time. The transmit power can be allocated at the AP side with total power constraint by splitting the power among all the subcarriers for different SUs. Thus, the optimal power control needs to jointly consider sensing and transmission in different subcarriers.
\end{itemize}

\subsection{Multiple Networks Coexistence}

Spectrum sharing has been recognized as a key remedy for the spectrum scarcity problem, especially after the successful deployment of WLAN and WPAN devices on an unlicensed band (e.g. ISM band). However, severe performance degradation has been observed when heterogeneous devices share the same frequency band due to mutual interference rooted in the lack of coordination. The cooperative busy tone~(CBT) algorithm allows a separate node to schedule a busy tone concurrently with the desired transmission, and thereby improving the visibility among difference sorts of devices~\cite{Zhang2011}. But preventing the busy signal from interfering the data packet still remains a problem. By deploying the FD techniques, the coexistence between heterogeneous networks may become more flexible. The research problem is to further reduce the RSI impact and realize efficient spectrum access management.

\section{Conclusions}

This article presented a new paradigm for future CR by exploring the FD technology to allow the SUs to simultaneously sense and access the vacant spectrum. Novel protocol design and key parameter derivation have been explained in depth. Both analytical and simulated results indicated that the proposed LAT protocol can efficiently improve the spectrum utilization. Feasible applications with FD enabled CR have been elaborated into centralized and distributed scenarios. The associated signal processing and spectrum access problems in these systems were also outlined. Future work may further study the scenario with multiple PUs and SUs in CRNs, and introduce basic economic theories as a tool to study and analyze FD-CRNs, such as dynamic spectrum access and spectrum trading problems.

\newpage
\begin{biography}{Yun Liao}
(S'14) has been pursuing her B.S. degree in School of Electrical Engineering and Computer Science in Peking University since 2011. Her research interests mainly include cognitive radio networks, full-duplex communications, cooperative communications, wireless network virtualization, and optimization theory.
\end{biography}

\begin{biography}{Lingyang Song}
(S'03-M'06-SM'12) received his PhD from the University of York, UK, in 2007, where he received the K. M. Stott Prize for excellent research. He worked as a research fellow at the University of Oslo, Norway, and Harvard University, until rejoining Philips Research UK in March 2008. In May 2009, he joined the School of Electronics Engineering and Computer Science, Peking University, China, as a full professor. His main research interests include MIMO, cognitive and cooperative communications, physical layer security, and wireless ad hoc/sensor networks. Dr. Song published extensively and wrote 3 text books. He is the recipient of 2012 IEEE Asia Pacific (AP) Young Researcher Award, and received 7 best paper awards, including the best paper award in IEEE International Conference on Wireless Communications, Networking and Mobile Computing (WCNM 2007), the best paper award in the First IEEE International Conference on Communications in China (ICCC 2012), the best student paper award in the 7th International Conference on Communications and Networking in China (ChinaCom2012), the best paper award in IEEE Wireless Communication and Networking Conference (WCNC2012), the best paper awards in International Conference on Wireless Communications and Signal Processing (WCSP 2012), the best paper awards in IEEE International Conference on Communications (ICC 2014), and the best paper awards in IEEE Global  Communication Conferences (Globecom 2014). Dr. Song is IEEE senior member and IEEE distinguished lecturer since 2015.
\end{biography}

\begin{biography}{Zhu Han}
(S'01-M'04-SM'09-F'14) received the B.S. degree in electronic engineering from Tsinghua University, in 1997, and the M.S. and Ph.D. degrees in electrical engineering from the University of Maryland, College Park, in 1999 and 2003, respectively.

From 2000 to 2002, he was an R$\&$D Engineer of JDSU, Germantown, Maryland. From 2003 to 2006, he was a Research Associate at the University of Maryland. From 2006 to 2008, he was an assistant professor in Boise State University, Idaho. Currently, he is an Associate Professor in Electrical and Computer Engineering Department at the University of Houston, Texas. His research interests include wireless resource allocation and management, wireless communications and networking, game theory, wireless multimedia, security, and smart grid communication. Dr. Han is an Associate Editor of IEEE Transactions on Wireless Communications since 2010. Dr. Han is the winner of IEEE Fred W. Ellersick Prize 2011. Dr. Han is an NSF CAREER award recipient 2010. Dr. Han is IEEE Distinguished lecturer since 2015.
\end{biography}

\begin{biography}{Yonghui Li}
(M'04-SM'09) received the Ph.D. degree from Beijing University of Aeronautics and Astronautics Beijing, China, in 2002. From 1999 to 2003, he was affiliated with Linkair Communication Inc, where he held a position of project manager with responsibility for the design of physical layer solutions for the LAS-CDMA system. Since 2003, he has been with the Centre of Excellence in Telecommunications, the University of Sydney, Sydney, N.S.W., Australia. He is now an Associate Professor in School of Electrical and Information Engineering, University of Sydney. He was the Australian Queen Elizabeth II Fellow and is currently the Australian Future Fellow. His current research interests are in the area of wireless communications, with a particular focus on MIMO, cooperative communications, coding techniques and wireless sensor networks. He holds a number of patents granted and pending in these fields. He is an executive editor for European Transactions on Telecommunications (ETT). He has also been involved in the technical committee of several international conferences, such as ICC, Globecom, etc.
\end{biography}

\end{document}